\documentclass[11pt]{article}
\usepackage{graphicx}
\begin{document}
\sloppy
\author{\textbf{Arbab I. Arbab}\\
\\
ICTP, P.O. Box 586, Trieste, Italy}
\title{Evolution of  angular momenta and energy of the Earth-Moon system}
\maketitle \begin{abstract} We have developed a model for the
evolution of the Earth-Moon angular momenta, energy dissipation
and tidal torque valid for the entire history of the Earth-Moon
system. The model is supported by present observational data.
\end{abstract}
\vspace{2cm} \textbf{Key Words}: {Earth-Moon system: tidal torque,
angular momentum, rotation, dissipation} \vspace{.5in}
\baselineskip=20pt
\section{Introduction}
Very recently, we have developed a model to calculate the length
of year, day, month and angular momentum of the Earth-Moon system
in the past time (Arbab, 2003 b). By considering the effect of
expansion of the universe on the Earth - Moon system, we were able
to explain the geologic evolution of the Earth's past rotation. We
have shown that the perturbation affecting our Earth-Moon system
can be modelled by an introducing an effective gravitational
constant which embodies these perturbations, so that we keep
Newton's law of gravitation and Kepler's third law invariant, and
applicable to our system. We have found that the this cosmic
evolution is manifested in several geologic phenomena (e.g. tides,
which are the result of gravitational forces, etc). Gravitational
forces in the Earth-Mon system cause tides, or bulge in the shape
Earth and the Moon.  It costs a lot of energy to deform the Earth,
and this energy is lost through the internal friction of rock
rubbing against rock within the Earth to raise the solid body
tides. Tides raised in the oceans by the Sun and Moon on Earth
dissipate significant energy, and transfer angular momentum from
the spin of the Earth to the orbit of the Moon. However, since
much of the dissipation occurs in shallow marginal seas, the
dissipation rate will not be constant. Glacial cycle changes in
sea level  and tectonic changes in coastal morphology will both
influence the evolution of the lunar orbit.

Various theories of the Earth's evolution have suggested that its
moment of inertia has gradually changed. While Lyttleton finds
that the Earth's moment of inertia has decreased by 30\% during
the last 3 billion years, others predict that it was smaller in
the past than at present. Changes in mass distribution can
significantly change the rotational dynamics of the Earth-Moon
system. Carey has suggested a doubling of the Earth's moment of
inertia ($C$) since the paleozoic. Any change in the moment of
inertia will induce a change in the rate of the Earth's rotation.
The transfer of angular momentum from Earth to Moon resulted in an
appreciable increase in the length of day and a decrease in the
length of month. Since the Earth's spinning rate is slowing down
the Moon's orbital angular momentum ($L$) has to change in order
to keep the total angular momentum of the system constant. It is
widely understood that an increase in the angular momentum of the
Moon immediately implies an increase in the Moon-Earth distance.
However, we have shown that is not always the case. In an
expanding universe one can satisfy this if the gravitational
constant is replaced with an effective constant involving a time
coordinate that takes care of the universe expansion, keeping the
normal Newton's constant invariant (Arbab, 2003a).

With the present Apollo Moon's retreat of 3.82 cm/y when
extrapolated backward would mean the Moon can not exist as a
stable body more than 1.5 billion years ago. The tidal force would
have torn it apart. It is also remarked by some astronomers and
geologists that the tidal force was smaller in the past than now.
Slichter (1963) remarked that if "for some unknown reason" the
tidal torque was much less in the past than in the present (where
"present" means roughly the last 100 million years), this would
solve the problem. But he could not supply the reason, and
concluded his paper by saying that the time scale of the
Earth-Moon system "still presents a major problem". Stacey (1977,
p.103) concluded that "...tidal friction was very much less in the
remote past than we would deduce on the basis of present-day
observations".So far geologists have not found any geologic
evidence of mega-tide to support a close approach of the Moon in
the past. The tidal evolution is critical for all theories of the
evolution of the Earth-Moon system. The schematic backward
integrations reaches a narrow state in too short a time into the
past, in contradiction with the lack of any geological evidence.
Older fossils show that tides existed as long as 2.8 billion years
ago and that the month was 17 days long (Kaula and Harris, 1975).
Our model however, gives a value of 19.6 days. Pannella (1972)
suggests that there may have been 39 days per synodic month 2
billion years ago. However, our model says there are 37.6 days per
synodic month.
\section{Earth-Moon angular momenta, angular velocities and torque}

Runcorn (1964) has shown that the value of the orbital angular
momentum of the Moon at present ($L_0$) to its value 370 million
years ago (since Devonian)
\begin{equation}\frac{L_0}{L}= 1.016\pm 0.003.
\end{equation}
He then found that the lunar tidal torque acting on Earth since
the Devonian to be  $3.9\times 0^{16}\rm\ N m$.\\ Conservation of
the angular momentum of the Earth-Moon system implies
\begin{equation}
L+S=L_{\rm tot}=\rm constant,
\end{equation}
where  $S$ is the spin angular momentum of the Earth. At present
one has
\begin{equation}
\frac{L_0}{S_0}=4.83
\end{equation}
Hereafter the subscript `0' denotes present day quantity.

We have found earlier (Arbab, 2003b) that the
\begin{equation}
 L=L_0\left(\frac{t_0-t}{t_0}\right)^{0.44},
\end{equation}
where $t$ is the time measured before present and $t_0=11\times
10^9 \rm\ y$ (the present age of the universe\footnote{See
Chaboyer, B. {\it et al.}, 1998. {\it Astrophys. J., 494}, 96.}).
With $L_0=2.878\times 10^{34}\rm\ kg\ m^2$ eq.(3) gives
$S_0=0.5958\times 10^{34}\rm\ kg\ m^2$ (see Touma and Wisdom,
1994) so that $L_{\rm tot}=3.4738\times 10^{34}\rm\ kg\ m^2$.\\
Sonett {\it et al.} (1996)
 have obtained a value of $2.727\times 10^{34}\rm J\ s$ for the orbital angular momentum of the Moon
whereas our value is $2.774\times 10^{34}\rm J\ s$.

The Moon mean orbital  motion is given by (Arbab, 2003b)
\begin{equation}
n=n_0\left(\frac{t_0-t}{t_0}\right)^{1.3}
\end{equation}
 and the deceleration rate of the Earth's spin
\begin{equation}
 \omega=\omega_0\left(\frac{t_0}{t_0-t}\right)^{2.6}.
\end{equation}
The above two equations yield the relation
\begin{equation}
 \omega\ n^2=\omega_0\ n_0^2\ ,
\end{equation}
which relates directly the Earth angular velocity to the Moon
angular velocity. Thus if the Earth deceleration rate is known,
the Moon acceleration rate can be found. The above equation yields
\begin{equation}
 \left(\frac{\dot\omega}{\omega}\right)=-2\left(\frac{\dot n}{n}\right)\ ,
\end{equation}
where the dot defines the derivative with respect to time.
Equations (5) and (6) show that the length of day and month were
respectively, 6 hours and 14 days (preset days) when the
Earth-Moon system formed. Thus, the month lengthens to 29.5 days
and the day to 24 hours!

 From eqs. (2) and (3) the moment of inertia becomes
\begin{equation}\label{}
C=\frac{L-L_{\rm tot}}{\omega}\ ,
\end{equation}
where $\omega_0=7.2921\times 10^{-5}\ \rm rad \ s^{-1},\
n_0=2.66170 \times 10^{-6}\ \rm rad \ s^{-1}$.   We remark that
during the time between 1000-3000 million years ago the Earth has
undergone a fast deceleration, and we attribute this to the
emergence of abundant water on Earth. We have noticed that the
moment of inertia has changed very slightly over the last 100
million years but has been doubled since the Earth formation.
According to our model, the present increase in the moment of
inertia is given by
\begin{equation}
\left(\frac{dC}{dt}\right)_0=3.9\times 10^{29}\rm\ kg\ m^2\
 cy^{-1}
\end{equation}
in comparison with a decreasing value obtained by  Bursa (1987) of
\begin{equation}
\left(\frac{dC}{dt}\right)_0=-(4.54\pm 0.49)\times 10^{29}\rm\ kg\
m^2\ cy^{-1},
\end{equation}
  where $C_0=8.17\times 10^{37}\rm\ kg\ m^2$. The values of $L, C, S $ and $\tau$ are
tabulated in Table 1 for different geologic epochs. Bursa (1986)
has evaluated the effects due to changes in the Earth's polar
moment of inertia due to the secular time variations in the second
zonal harmonic:
\begin{equation}
\frac{dC}{dt}=-\frac{2}{3}MR^2\frac{dJ_2}{dt} \end{equation} where
$MR^2=2.4296\times 10^{38}\rm kg\ m^2$ ($R=\rm Earth's\ \ radius,\
M=\rm Earth's \ \ mass$). Yoder {\it et al.} (1983) have estimated
that
\begin{equation}
\left(\frac{dJ_2}{dt}\right)_0=(-2.8\pm 0.6)\times 10^{-9}/\rm cy
\end{equation}
from LAGEOS tracking data and attribute it to the effects of
postglacial rebound. \\ Our model, however, gives
\begin{equation}
\left(\frac{dJ_2}{dt}\right)_0=-2.4\times 10^{-9}/\rm cy
\end{equation}
showing a good agreement with observational data. \\
 According to this model, the lunar tidal torque ($\tau)$ is given
by
\begin{equation}
\tau=\frac{dL}{dt}=-\frac{dS}{dt} \ ,\end{equation} or
\begin{equation}
 \tau=\tau_0\left(\frac{t_0}{t_0-t}\right)^{0.56},
\end{equation}
where $\tau_0=3.64764\times 10^{16}\rm N\ m$ is the present tidal
torque. The torque acting on Earth since the Devonian is thus
$3.83\times 10^{16}\rm\ N\ m$, which agrees with Runcorn (1964)
result. It also agrees with that given from the measurements of
the longitudes of the Sun and Moon (Murray, 1957). Kolenkiewicz
{\it et al.} (1973) obtained a torque of ($4.8\pm 0.8)\times
10^{16}\rm N\ m$.  It is found recently by Sonett \emph{et al.}
(1996) that ($900\pm 100)$ million years ago the torque was
$9.6\times 10^{16}\rm\ N\ m$ estimated from a study of laminated
tidal sediments. However, our model gives a value of $3.826\times
10^{16}\rm N\ m$. In particular, the torque has decreased by 26 \%
since the Earth formed. Holmberg (1952) obtained  a torque of
$3.7\times 10^{15}\rm\ N\ m$ and Munk and MacDonald (1960) gave a
torque of $3\times 10^{15}\rm\ N\ m$. While the Earth's spin has
decreased by 50 \%, the Moon's orbital momentum has increased by
25 \% since their formation (4.5 billion years ago). The slow
decrease of the Earth's spin rate is partly due to an increase in
the Earth's moment of inertia in addition to a decrease in the
Earth's angular velocity.\\  One must consider the possibility
that small relative movements due either to tidal forces or to
mechanical stresses on the surface of the Earth may have taken
place between the different plates. The Earth's moment of inertia
is changed periodically by tides raised by the Sun and the Moon,
which distort the shape of the Earth. The moment of inertia can
also change due to changes in the Earth's internal temperature. A
decline in the Earth's spin velocity would lead to an increase in
the polar diameter and a decrease in the equatorial diameter. Any
redistribution of mass in the Earth's interior would affect the
spinning rate of the Earth. This mass redistribution is mainly due
to tidal force that acing on Earth since its existence. It
dictates the Earth to maintain its equilibrium condition and in
turns the Earth exhibits some geological disturbances (e.g.,
volcanos, earthquakes, etc) in order to comply with the ever
changing conditions. Therefore, the Earth must have been suffering
a lot from the continuous geological disturbances in order to
remain in equilibrium. If tidal forces were bigger in the past
then its consequences on paleo-rocks must have been very
prominent. High tides would impose a big energy dissipation on
Earth in the past. However, Precambrian geology shows no global
melting could have occurred during the last 2.7 billion years, and
probably much earlier than that. We observe that the conservation
of angular momentum of the Earth-Moon system is responsible for
the increasing moment of inertia of the Earth. In this work we
analyze the evolution of the Earth-Moon system in regard to
angular momentum, energy and their dissipation. We came up with
excellent agreement with observations. These results encourage us
to look for other astronomical and geophysical application of this
model.
\section{Earth-Moon energy and tidal dissipation}
The Earth's rotational energy is given by
\begin{equation}
E_1=\frac{1}{2}C\omega^2,
\end{equation}
and the Moon's orbital energy is given by
\begin{equation}
E_2=-\frac{GmM}{2r},
\end{equation}
where $r$ , $m$ and $M$ are the Earth-Moon distance, Moon's mass
and Earth's mass, respectively, and $C$ is given by eq.(9). \\
The rate of total tidal energy($E=E_1+E_2$) dissipated in the
Earth-Moon system is given by
\begin{equation}
P=-\frac{dE}{dt}\ .
\end{equation}
Recently, Sonett {\it et al.} (1996) have found that the orbital
energy of the Moon 900 million years ago to be $-4.242\times
10^{28}\rm J$. This coincides with our model prediction, which is
$-4.124\times 10^{28}\rm J$. MacDonald (1964) estimated an energy
of $1.5\times 10^{31}\rm \ J$ to be released on an overall slowing
down of the rotation of the Earth from a 3 hour to 24 hour period.
Monin {\it et al.} (1987) estimated that 4.0-3.2 billion years ago
the Earth released a tidal energy of $8.5\times 10^{29}\rm \ J$,
which was sufficient to melt the upper mantle to a depth of
350-400 km and which does not differ largely from our anticipated
value of $7.8\times 10^{29}\rm \ J$. Our model shows that the
total energy lost by the Earth-Moon system over the past 4.5
billion years was $1.48\times 10^{30}\rm \ J$. This enormous
energy must have been distributed (utilized) somewhere by the
Earth'body or oceans. Such a problem can be better tackled by
invoking geophysical and oceanographical treatment. We,  however,
remark that the MacDonald's value above is two order of magnitude
higher than observations might suggest.
\\
Wunsch (2000) estimated that the total presently observed energy
dissipation is about $3\times 10^{12}\rm W$. Egbert and Ray (2000)
reported from a TOPEX/Poseidon satellite that about $1\times
10^{12}\rm W$ (25-30\% of the total tidal energy dissipation)
occurs in the deep ocean. While the Earth has lost about 87\% of
its  original rotational energy, the Moon  lost only 21 \% of its
original orbital energy. The Earth present rotational energy is
$2.17\times 10^{29}\rm J$ which was  $1.687\times 10^{30}\rm J$
some 4.5 billion years ago. Thus early in the Earth history the
orbital source of power generation may have competed with
radiogenic one. This rotational source of energy may have powered
core convection and dynamo though they operate today at a smaller
rate. They are still strong to be detected by present lunar laser
techniques.

The astronomical estimate of the amount of tidal energy dissipated
in the Earth-Moon system
 is given by (Lang, 1992; Lambeck, 1980)
\begin{equation}
P=-4\times 10^{12}\rm \ W\ ,
\end{equation}
to be compared with the one calculated from the present recession
rate of $3.82\rm \ cm/y$, deceleration rate,  and with constant
moment of inertia, which is
\begin{equation}
P=-2\times 10^{12}\rm \ W\ .
\end{equation}
It is also suggested by Jeffreys (1976) that most reliable values
indicate that $P=-2.9\times 10^{12}\rm\ W$, and is twice the value
he obtained previously; and compared with astronomical values
$P=-2.7\times 10^{12}\rm\ W$ and $P=-0.6\times 10^{12}\rm\ W$
(Jeffreys, 1976). The  analysis of Hendershott (1972) favored a
value between 3 and  $4\times 10^{12}\rm\ W$. Munk and MacDonald
obtained (1960) a value of $P=-3.2\times 10^{12}\rm\ W$. Smith and
Jungles (1970) obtain a 3 to $5\times 10^{12}\rm\ W$ from
observational considerations of gravitationally determined tidal
phase lag. According to our model, we obtain a very close value of
\begin{equation}
P=-2.98\times 10^{12}\rm \ W\ ,
\end{equation}
compared with  the observed value in eq.(18). This coincides
exactly with the result obtained by Kagan and Kivman (1995) for
the $M_2$ ocean tide. The same number is also found by Pekeris and
Accad (1969). It is believed  that oceans provide the predominant
sink for the tidal energy: dissipation in the solid Earth and Moon
can be at most about 10\% of the total. Our model, however gives
the total contribution of the energy dissipation without giving
the details. (Stacey \& Stacey, 1999) have found that the total
tidal dissipation in 4.5 billion years to be about $2\times
10^{30}\rm\ J$ in comparison with our model's value of $1.47\times
10^{30}\rm\ J$. This shows how close our results are compatible
with present data. Fig.1 - Fig.4 show a graphical variation
$\omega$, $L$, $\tau$ and $E$ as functions of time. Our model will
benefit very from the future space and geophysical results.

As to date there is no complete account of the Earth history
related to its rotation, this model would satisfactory fill the
gap. We remark that it is the first simple model that relates the
Earth-Moon system parameters to time directly and explicitly.
\section{Acknowledgments} I would like to thank the abdus salam International Center for Theoretical Physics
for hospitality and Comboni College for providing financial
support. I wish to thank Prof. Heiko P\"{a}like (Cambridge) for
useful communication.
\section*{References}
Arbab, A.I., \emph{to appear in} {\it Acta Geod. Geoph. Hung.}, (2003a).\\
Arbab, A.I., {\it Class. Quantum Gravit.} \textbf{20}, 93 (2003b).\\
Bursa, M., {\it 26$^{th}$  COSPAR conference, Cent. Natl. d'Etudes
Spatiales}, Toulouse, France, (1986).\\
Bursa, M., {\it Bull. Astron. Inst. Czecgosl.}, \textbf{38}, 321 (1987).\\
Carey, S.W., {\it Continental Drift} (Univ. of Tasmania, Hobard
1958).\\
Egbert, G. D., and Ray, R. D., {\it Nature} \textbf{405}, 775 (2000).\\
Jeffreys, H. {\it The Earth: Its history and physical
constitution}, CUP, (1976).\\
Hendershott, M.C., {\it Geophys. J}, \textbf{29}, 389 (1972).\\
Holmberg, R. R., {\it Mon. Not. Roy. Astron.Soc. Geophys. Supp.} 6, 325 (1952). \\
Kaula, W.M., and Harris, A.W., {\it Geophys. Space Phys.},
\textbf{13}, 363
(1975).\\
Kagan, B.A., and Kivman, G.A., {\it Acta Geodesy and Geophysics
Hungary}, \textbf{80}, 135 (1995).\\
Kolenkiewicz, R., Smith, D.E., and Dunn, P.J., {\it Am. Geophys.
Un.}, \textbf{54}, 232 (1973).\\
Lambeck, K., {\it The Earth variable rotation, pp. 319}, CUP,
(1980).\\
Lang, K. R.,  \emph{Astrophysical Data: Planets and Stars}, New
York: Springer-Verlag, (1992).\\
Lyttleton, R.A., {\it Proc. Roy. Soc.}, \textbf{A275}, 1 (1963).\\
MacDonald, J.F., {\it Rev. Geophys.}, \textbf{2}, 467 (1964).\\
Monin, A.S., Sorokhtin, O.G., and Ushakov, S.A., {\it Akaeemiia
Nauk SSSR, Doklady}, \textbf{293}, 1341 (1987).\\
Munk, W.H., and MacDonald, G.J.F., {\it The rotation of the Earth},  CUP (1960).\\
Murray, C., {\it Mon. Not. Roy.Astro.}, \textbf{117}, 478 (1957).\\
Pannella, G.,  {\it Astrophys. Space Sci. 16}, 212 (1972).\\
Pekeris, C.L., and Accad, Y., {\it Phil. Roy. Soc. London.},
\textbf{A} 264, 413 (1969).\\
Runcorn, S.K, {\it Nature}, \textbf{204}, 824 (1963).\\
Slichter, Louis B., {\it Journal of Geophysical Research}, 68, 14,
(1963).\\
Smith, S.W., and Jungles, P., {\it Physics of Earth and Planet
Interiors}, \textbf{2}, 233 (1970).\\
Sonett, C.P., Kvale, E.P., Zakharian, A., Chan, M.A., and Demko,
T.M., {\it Science} \textbf{273}, 100 (1996).\\
Stacey, F.D., {\it Physics of the Earth}, John Wiley \& Sons. Inc.
(1977).\\
Stacey, F.D., and Stacey, C.H.B., {\it Phy,  Earth. Plan.
Interior}, \textbf{110}, 83 (1999).\\
Touma, J., and Wisdom, J., {\it Astronomical Journal},
\textbf{108}, 1943 (1994).\\
Wunsch, C., {\it Nature} \textbf{405}, 743 (2000).\\
Yoder, C. F., {\it  et al.}, {\it Nature}, \textbf{303}, 757 (1983).\\
\newpage
\begin{table}
\caption{The Earth-Moon system parameters at different geologic
times} \vspace{1cm}
\begin{tabular}{r|r|r|r|r|r|r|r|r|r}
\hline
Time $^*$  & Now & 65 & 136 &  180  & 230 & 280  & 345  & 405 & 500 \\
\hline
$L$ ($\times 10^{34}\rm\ kg\ m^2$)& 2.878 & 2.871 & 2.863 & 2.858 & 2.852 & 2.846 & 2.839 & 2.832 & 2.821 \\
$I$ ($\times 10^{37}\rm\ kg\ m^2$) & 8.174 & 8.144 & 8.114 & 8.095 & 8.072 & 8.048 & 8.015 & 7.984 & 7.932 \\
$\tau$ ($\times 10^{16}\rm\ N\ m$)& 3.648 & 3.659 & 3.673 & 3.681 & 3.691 & 3.700 & 3.713 & 3.725 & 3.744 \\
$\omega$ ($\times 10^{-5}\rm\ rad\ s^{-1}$)& 7.292 & 7.405 & 7.530 &  7.611 & 7.703 & 7.797 & 7.922 & 8.039 & 8.229  \\
$n$ ($\times 10^{-6   }\rm\ rad\ s^{-1}$)& 2.6617 & 2.64127 & 2.61936 & 2.60522 & 2.58958 & 2.75396 & 2.55369 & 2.53501 & 2.50550\\
$E$ ($\times 10^{28}\rm\ J$)& 17.44 & 18.06 & 18.75 & 19.20 & 19.71 & 20.23 & 20.93 & 21.59 & 22.66\\
\hline\hline
\end{tabular}
\end{table}
\vspace{2cm}
\begin{table}
\begin{tabular}{r|r|r|r|r|r|r|r|r|r}
\hline
Time $^*$  & 600 & 900 &1000 & 1300 & 1750 & 2170 & 3000 & 3500 & 4500\\
\hline
$L$ ($\times 10^{34}\rm\ kg\ m^2$)&  2.809 &  2.774 & 2.762 & 2.726 & 2.671 & 2.619 & 2.510 & 2.441 & 2.295 \\
$I$ ($\times 10^{37}\rm\ kg\ m^2$) & 7.875 & 7.684 & 7.614 & 7.390 & 7.013 & 6.624 & 5.777 & 5.232 & 4.115  \\
$\tau$ ($\times 10^{16}\rm\ N\ m$)& 3.764 & 3.826 & 3.847 & 3.913 & 4.019 & 4.125 &   4.359 &4.520 & 4.897 \\
$\omega$ ($\times 10^{-5}\rm\ rad\ s^{-1}$)& 8.437 & 9.1004 & 9.342 & 10.112 & 11.442 & 12.911 & 16.689 & 19.738 & 28.365\\
$n$ ($\times 10^{-6}\rm\ rad\ s^{-1}$)& 2.47453 & 2.38213 & 2.35152 & 2.26023 & 2.12487 & 2.00031 & 1.75940 & 1.61781 & 1.34319\\
$E$ ($\times 10^{28}\rm\ J$)& 23.85 & 27.72 & 29.12 & 33.73 & 41.94 & 51.33 & 76.73 & 98.31 & 165.33\\
\hline \hline
\end{tabular}
\end{table}
\emph{.}\\
$^*$ Time: In million years before present,\\
$S$= Spin angular momentum,\\
$L$= Orbital angular momentum,\\
$I$= Moment of Inertia,\\
$\tau=$ Torque.\\
$E$= Total Earth-Moon system energy
\newpage
\includegraphics [width=\textwidth , bb= 50 20 550 550]{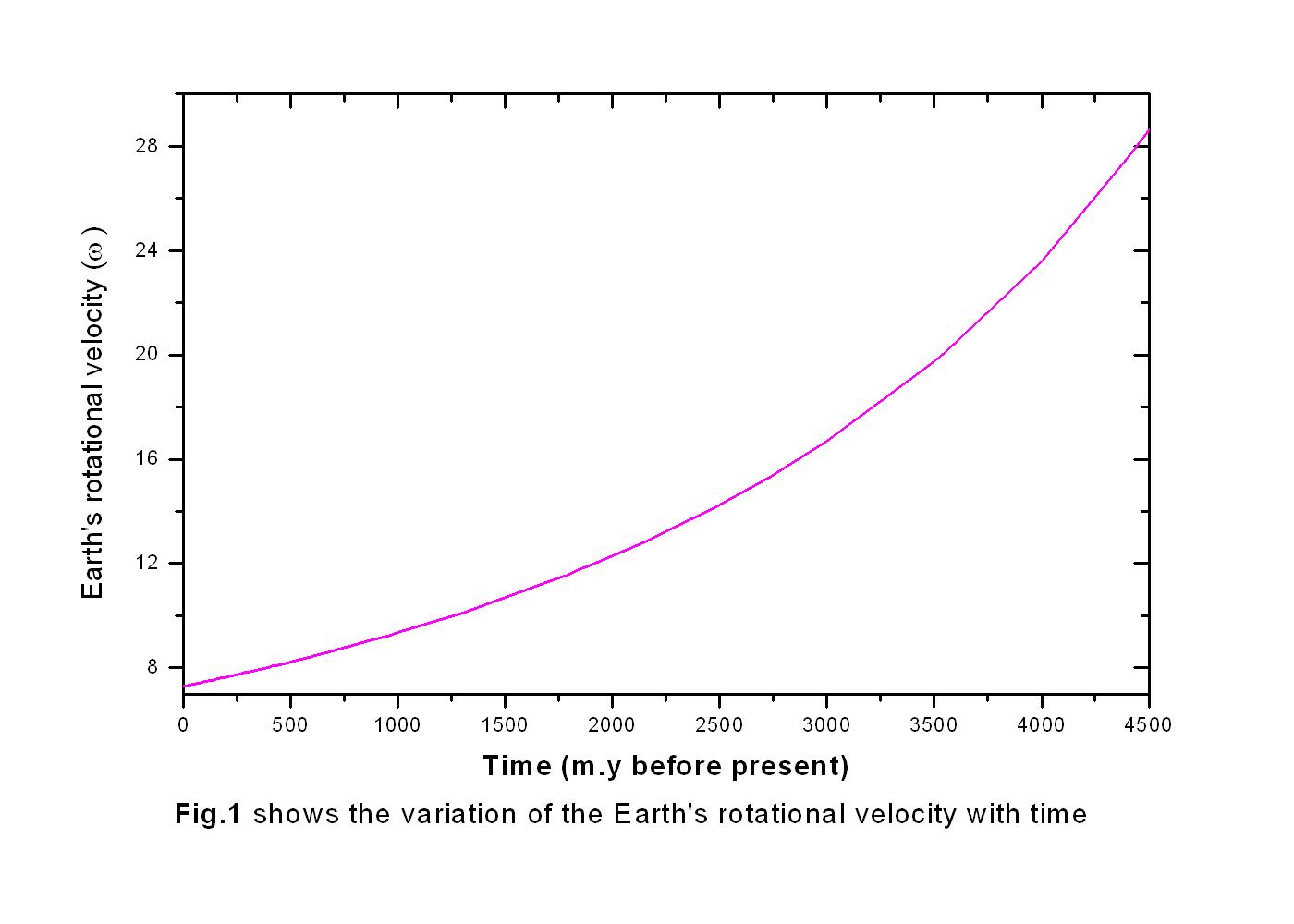}
\newpage
\includegraphics [width=\textwidth , bb= 50 20 550 550]{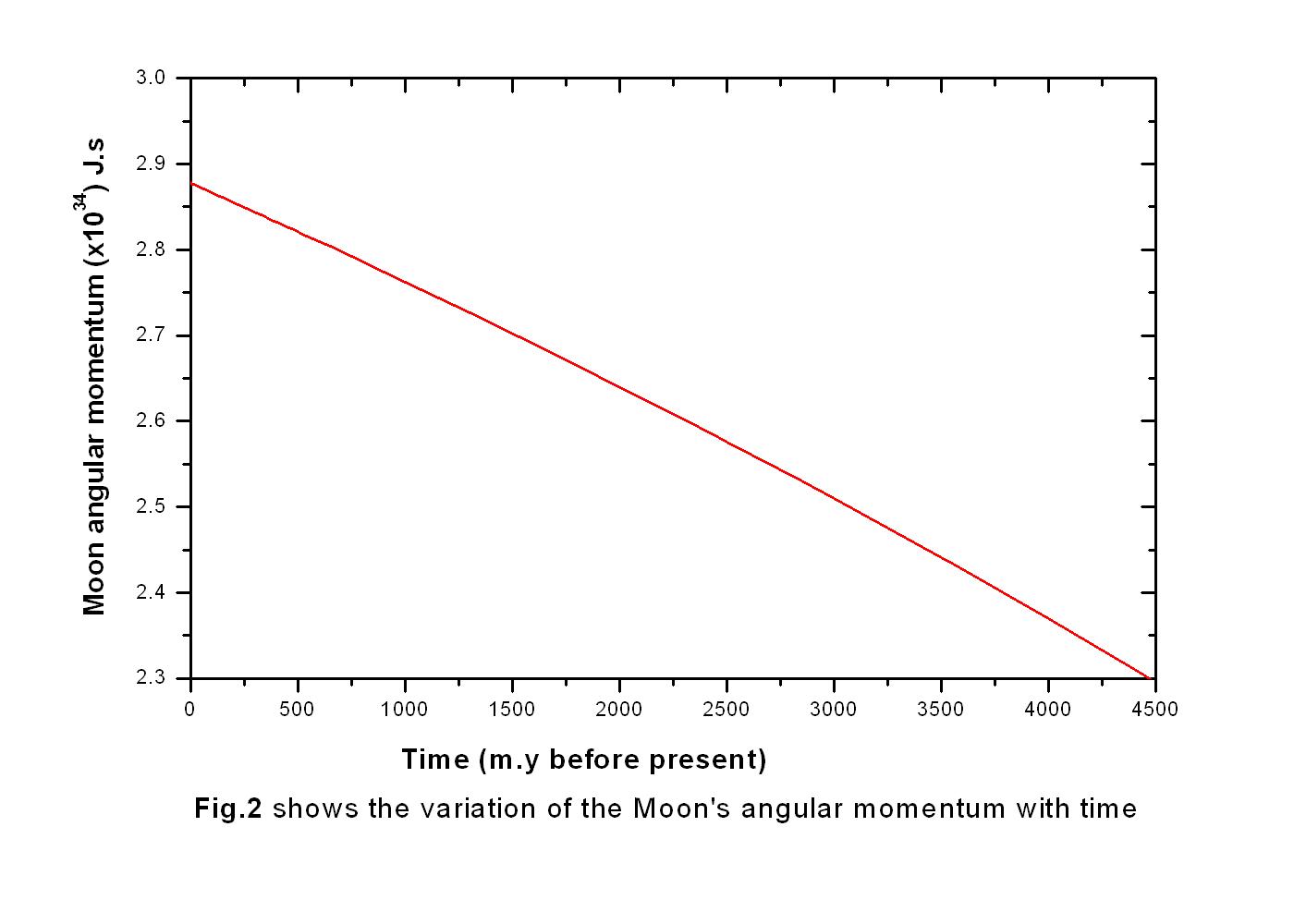}
\newpage
\includegraphics [width=\textwidth , bb= 50 20 550 550]{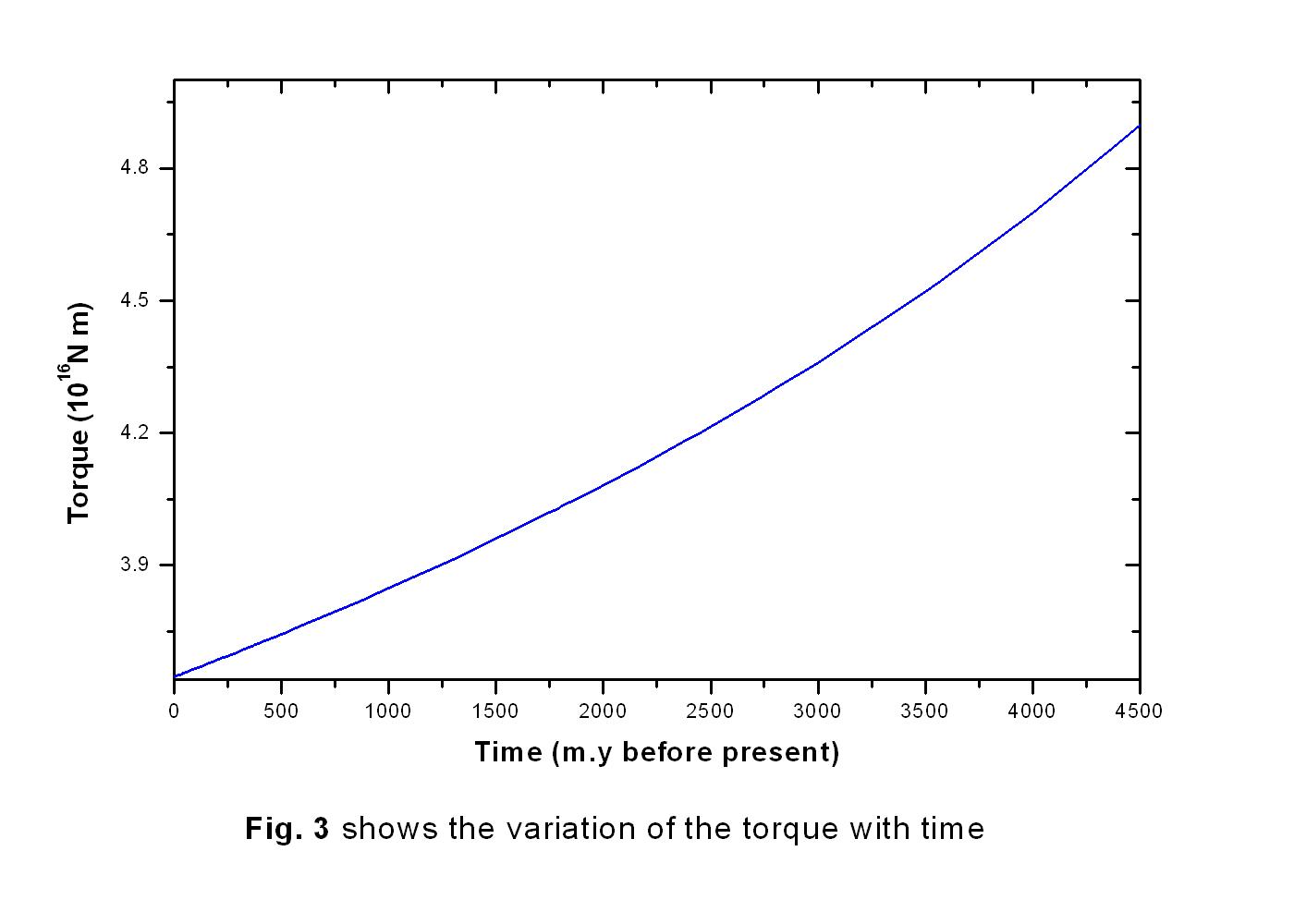}
\newpage
\includegraphics [width=\textwidth , bb= 50 20 550 550]{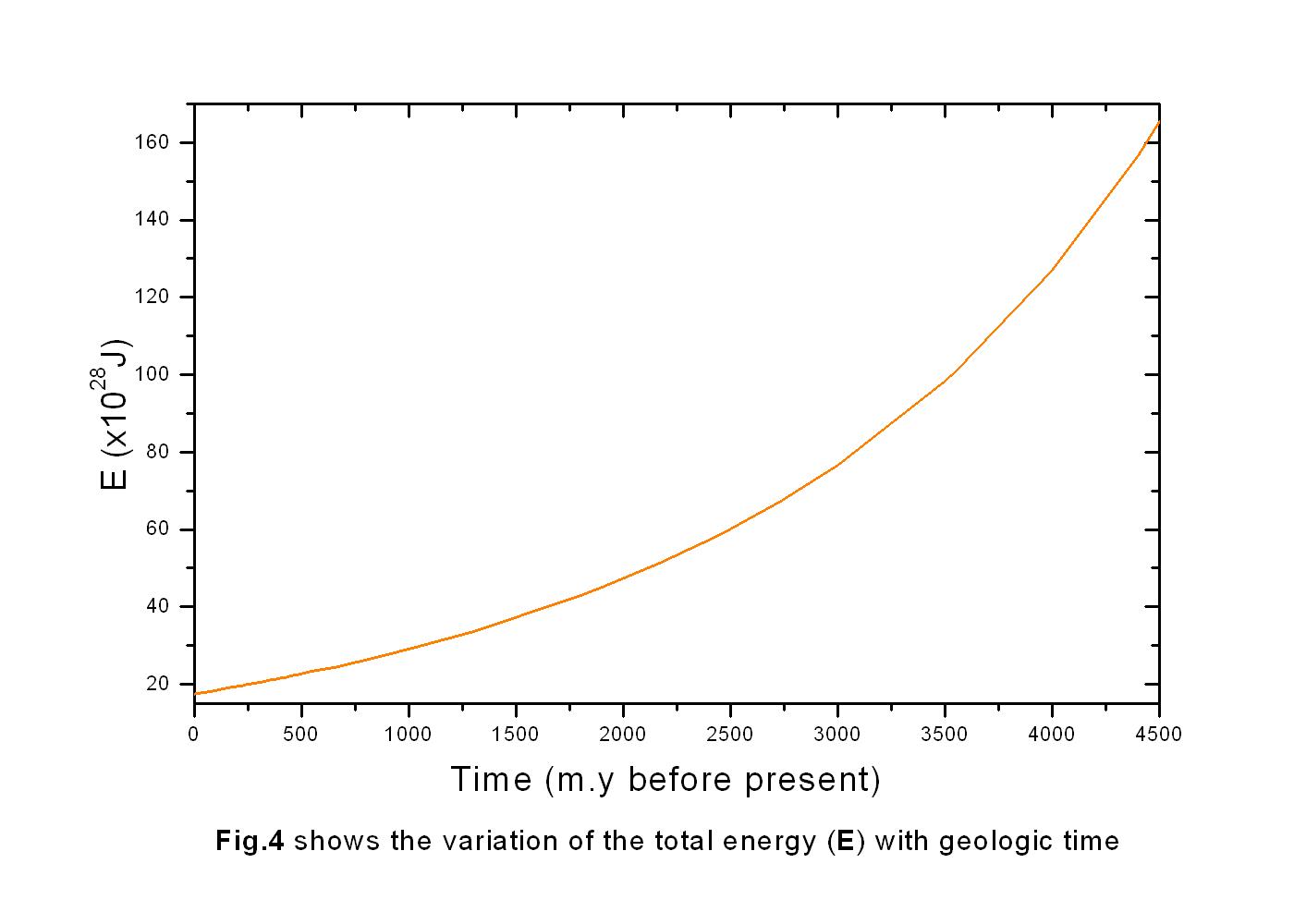}
\end{document}